\documentclass[11pt,a4paper]{article}
\usepackage[a4paper,margin=2.2cm]{geometry}
\usepackage{amsmath,amssymb}
\usepackage{bm}
\usepackage{graphicx}
\usepackage{caption}
\usepackage{subcaption}
\usepackage{booktabs,tabularx,array,longtable}
\usepackage{enumitem}
\usepackage{xcolor}
\usepackage{cite}
\usepackage{hyperref}
\usepackage{float}
\usepackage{microtype}
\usepackage{tikz}
\usetikzlibrary{shapes.geometric} 
\usepackage{titlesec}
\usepackage[none]{hyphenat} 
\usepackage{ragged2e}
\usepackage{xurl}

\usepackage[T1]{fontenc}
\usepackage{lmodern}
\usepackage{tabularx}
\usepackage{placeins}

\hypersetup{colorlinks=true, linkcolor=black, citecolor=black, urlcolor=blue}
\graphicspath{{figures/}{work/figures/}}
\titleformat{\section}{\Large\bfseries\sffamily\color{black}}{\thesection}{0.65em}{}
\titleformat{\subsection}{\large\bfseries\sffamily\color{black}}{\thesubsection}{0.65em}{}
\setlist[itemize]{leftmargin=1.5em,itemsep=2pt,topsep=2pt}
\newcolumntype{Y}{>{\raggedright\arraybackslash}X}

\begin{document}

\begin{center}

{\LARGE\bfseries Beyond Peak TOPS/W: A System-Level Perspective on Hybrid Digital, Analogue and Neuromorphic Computing\par}

\vspace{0.7em}
{\large\bfseries Eiman Kanjo\par}
{\em Computer Science, Nottingham Trent University; Department of Computing, Imperial College London\par}
eiman.kanjo@ntu.ac.uk; e.kanjo@imperial.ac.uk\par
{\large\bfseries Varuna De Silva\par}
{\em School of Computer Science and Digital Technologies, Aston University\par}
v.desilva@aston.ac.uk\par
\end{center}
\begin{abstract}
The digital revolution, which progressively replaced analogue methods with digital circuits, has entered a new phase as AI expands across cloud infrastructure, mobile networks, wearables and physical systems, including drones and robots. Digital computing remains the general-purpose foundation of this expansion:
it supports heterogeneous, on-device and decentralised AI through programmable control, mature software and decades of accumulated engineering infrastructure. Yet as energy and data-movement constraints become more significant, that same foundation is increasingly being extended rather than replaced by selected analogue and physical principles that it can host, configure and verify. Photonic, in-memory and neuromorphic architectures offer routes to reducing data movement and accelerating matrix-intensive and event-driven processing, not as alternatives to digital infrastructure but as specialised engines operating within it. This paper argues that hybrid digital--analogue computing represents a credible pathway towards more energy-efficient AI systems: one in which physical substrates earn an expanding role only where they deliver a measurable system-level advantage, under digital orchestration that manages integration, uncertainty and fallback. It examines the architectural principles, workload suitability, energy accounting, software requirements, limitations and open challenges associated with this transition, and argues that future progress
should be evaluated through deployed-system metrics rather than isolated peak
tera operations per second per watt (TOPS/W) claims.
\end{abstract}


\section{Motivation and central proposition}

AI is increasingly moving beyond data centres into mobile devices, embedded sensors, autonomous machines and distributed physical systems. These platforms operate under constraints that differ  from those of conventional cloud computing, including limited energy, memory and communication capacity, intermittent connectivity, environmental variation and real-time response requirements. Consequently,
energy efficiency cannot be treated solely as a property of an AI model or
computing chip. It emerges from interactions across sensing, computation,
memory movement, communication, software, hardware and, in autonomous
systems, actuation.

Digital computing has met much of this challenge on its own terms, through
model compression, quantisation, pruning, knowledge distillation,
specialised accelerators and memory-aware software~\cite{hinton2015distilling,gou2021knowledge,dally2020domain}.
These advances provide a strong and continually improving baseline, and
nothing in this paper should be read as suggesting otherwise: digital
computing's programmability, security, mature tooling and manufacturing
maturity remain unmatched, and any physical substrate discussed in what
follows is judged against that baseline rather than instead of it. Even so,
data movement between memory and processing units remains a major source of
energy consumption, particularly for matrix-intensive AI
workloads~\cite{horowitz2014computing}, and this specific limitation, rather
than any general shortfall in digital computing, is what has renewed interest in physical computing approaches that perform selected operations through the properties of devices and materials. Analogue in-memory arrays can combine weight storage with matrix operations; photonic systems can exploit optical propagation and interference; and neuromorphic architectures can process sparse temporal events through localised, event-driven computation.

These technologies should not be viewed as universal replacements for digital computing, and this paper does not treat them as such. Their practical advantages depend on workload structure, numerical precision, interface overhead, calibration, environmental stability and the maturity of supporting software, all of which digital orchestration exists precisely to
manage. A specialised physical accelerator may demonstrate high efficiency at the core while providing little or no advantage once data conversion, control, memory access and error correction are included; a carefully matched physical substrate, hosted and verified by a digital system built to
accommodate it, may instead reduce system-level energy or latency when assigned operations that align with its physical properties. The difference
between these two outcomes is not a property of the physical substrate alone, but of how well it has been integrated into a digital system designed to make that integration safe.

This perspective therefore examines the current state and recent advances in digital, analogue in-memory, photonic and neuromorphic computing, not as four separate trajectories but as elements of a single, increasingly integrated computing landscape. It considers the capabilities, maturity, limitations and suitable workloads of each paradigm, and identifies how they can operate as complementary parts of a heterogeneous system built on a digital foundation. 
The central proposition is that hybrid architectures are valuable when physical substrates deliver measurable end-to-end benefits relative to a strong digital baseline for well-matched workloads, rather than acting as complete general-purpose replacements.
Such benefits should be evaluated using deployed-system metrics rather than isolated accelerator performance. This account is written to be approachable to readers with a digital AI or software background as well as those from devices and materials, since progress on this front depends on both communities being able to follow, and eventually contribute to, the other's reasoning.

\section{Digital and Physical Computing Foundations}
\subsection{Digital computing}
\begin{figure}[htbp]
\centering
\includegraphics[width=0.80\linewidth]{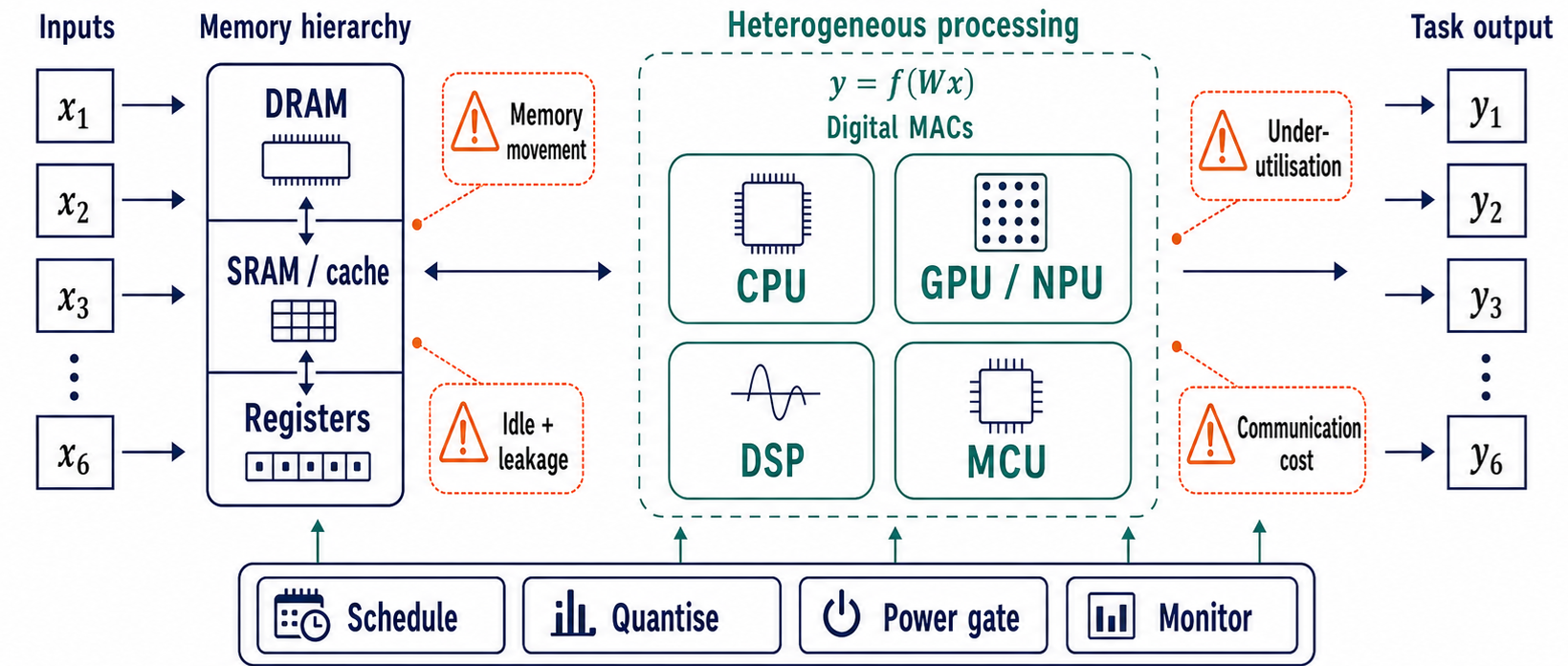}
\caption{Digital computing as the system-level baseline.}
\label{fig:digital-computing}
\end{figure}

Modern digital devices are increasingly heterogeneous. Mobile, embedded and edge platforms commonly combine CPUs, GPUs, neural processing units (NPUs), digital signal processors (DSPs), microcontrollers, memory controllers, sensor interfaces, wireless modems and power-management units. This integration allows workloads to be distributed across components with different performance, precision and energy characteristics. As illustrated in Figure~\ref{fig:digital-computing}, similar architectures are appearing in physical AI systems, including robots, drones and industrial monitoring platforms.
This development has been supported by advances in semiconductor integration, memory hierarchies, domain-specific accelerators, model compression and software  optimisation~\cite{dally2020domain,sze2017efficient,han2016deep}. Digital computing therefore remains the principal baseline for energy-efficient AI. It provides programmable and deterministic operation, mature software and security tools, flexible numerical precision and established manufacturing processes. Any proposed analogue, photonic or neuromorphic system should consequently be compared with an optimised digital implementation rather than a general-purpose or outdated baseline.

\subsubsection{Smaller and More Efficient Models}
The movement of AI towards local devices has been supported by methods that reduce model size, computation and memory traffic. These include pruning, quantisation, knowledge distillation, sparsity and hardware-aware model design~\cite{hinton2015distilling,han2016deep,frantar2023sparsegpt}. For language models, low-rank adaptation and quantisation-aware fine-tuning can reduce the cost of adapting and storing model parameters. Quantisation Aware - Low Rank Adaptation (QA-LoRA), for example, combines low-rank adaptation with quantisation-aware training to support low-bit deployment~\cite{xu2023qa}. T-MAC reformulates low-bit matrix multiplication as table-lookup operations and demonstrates how software–hardware co-design can improve language-model inference on CPUs~\cite{wei2024mac}.
These examples show that model efficiency is inseparable from execution architecture. A smaller parameter count does not necessarily produce a proportional reduction in energy or latency. The realised benefit depends on numerical representation, memory access, operator support, scheduling and hardware utilisation.

\textbf{On-Device Processing} 
Neural inference can now run on platforms with hundreds of kilobytes of SRAM and approximately one to two megabytes of flash~\cite{pullini2019mr}. Micro Controller Unit (MCU) based Network Search (MCUNet) jointly designs the neural architecture and inference runtime for microcontrollers, demonstrating ImageNet-scale inference on commercial MCUs under constrained SRAM and flash budgets~\cite{lin2020mcunet}. Memory-aware runtimes reduce tensor lifetimes, optimise operator scheduling and reuse activation buffers, enabling convolutional neural networks to fit within microcontroller-class memory budgets~\cite{lin2020mcunet,lin2021mcunetv}. This makes it possible to deploy tasks such as gesture recognition and acoustic or visual anomaly detection on embedded platforms such as Arm Cortex-M MCUs, Arduino-class boards and other low-power sensor nodes~\cite{gibbs2023combining,ren2021tinyol}.

The transition to edge AI provides low latency, enhanced privacy and reduced transmission costs while mitigating unreliable connectivity in remote or contested environments. Sustaining long-term field deployment demands strict, system-level management of active power. For example, in agriculture, devices may be deployed for long periods with limited maintenance, requiring close hardware--software integration to process multimodal streams directly on-device~\cite{patrick2026review,zhang2025multicore}. Power management is therefore a system-level design requirement rather than a secondary implementation detail~\cite{parisi2019continual,woodward2026hybrid}.

\subsubsection{Power Management and Event-Driven Operation}
Power management is now as important as model accuracy for many deployed systems. Efficient operation depends on active power, sleep current, sensing duty cycle, radio transmission and thermal constraints. In remote locations where internet infrastructure is absent, autonomous sensor nodes must orchestrate localised tracking and anomaly alerts without relying on cloud computation~\cite{patrick2024internetless}. Adaptive, decentralised framework topologies enable these resource-constrained nodes to collaborate efficiently within strict hardware envelopes~\cite{kanjo2026node}. Event-driven sensing methods, including wake-on-sound, wake-on-motion and wake-on-vision, reduce unnecessary processing. Dynamic voltage and frequency scaling (DVFS), power gating and duty-cycled communication reduce energy use when the device is idle or when only low-complexity monitoring is needed.
Research on energy-harvesting TinyML demonstrates the value of combining model compression, low-energy execution and intermittent-computing support \cite{islam2021enabling}. 
\subsubsection{Adaptive and Collaborative Intelligence}
Deployed models may encounter sensor drift, changing environments, new users and evolving operational requirements. Continual learning and selective model updates can allow systems to adapt without complete retraining~\cite{parisi2019continual}. TinyOL demonstrates online learning from streaming data on constrained microcontrollers~\cite{ren2021tinyol}. Such adaptation introduces additional energy, memory, reliability and validation costs that should be included in system-level evaluation.
When multiple devices collaborate, the system must also account for communication, synchronisation and hardware heterogeneity. Conventional federated learning coordinates local training through a central aggregation process~\cite{mcmahan2017communication}. Node Learning considers a more decentralised setting in which individual nodes maintain local models and exchange knowledge opportunistically without requiring a permanent central aggregator or uniform hardware~\cite{kanjo2026node}. These approaches extend the system boundary from an individual accelerator to a network of sensing, computing and communicating devices.

\subsection{Analogue and Physical Computing}
Analogue and physical computing use the behaviour of devices, materials or waves to perform selected mathematical operations. Rather than representing every operation as a sequence of digital instructions, these systems map computation onto physical quantities such as electrical conductance, current, phase or optical intensity.  This can provide parallelism and reduce some forms of data movement when the physical operation closely matches the computational workload

For AI, the main opportunity lies in operations that are repeated extensively, particularly matrix--vector multiplication and the processing of continuous or sparse sensor signals. The physical operation is rarely sufficient on its own. Input encoding, output conversion, nonlinear functions, memory management, calibration and system control generally remain digital. The resulting architecture is therefore usually mixed-signal or hybrid rather than purely analogue.

This perspective focuses on two physical-computing approaches:

\begin{itemize}
    \item \textbf{Analogue in-memory computing:} programmable conductance
    arrays store weights and perform matrix operations through electrical current accumulation~\cite{sebastian2020memory,gallo2018mixed}.
    
    \item \textbf{Photonic computing:} optical interference, propagation and
    wavelength multiplexing are used to implement high-bandwidth linear transformations~\cite{clements2016optimal}.
\end{itemize}

\subsubsection{Analogue In-Memory Computing}
In analogue in-memory computing (AIMC), matrix weights are represented by programmable conductance values within a crossbar array~\cite{sebastian2020memory,
gallo2018mixed}. At the intersection of each row and column is a programmable
memory device or circuit that represents the weight element \(G_{ij}\). Many
proposed non-volatile implementations use memristive devices, including resistive
random-access memory (RRAM) and phase-change memory (PCM)~\cite{zidan2018future}.
These devices retain their programmed conductance states without a continuous
power supply, although peripheral circuitry and control logic still consume
energy during system operation.
An individual device commonly represents a non-negative conductance. Signed
weights may therefore require differential device pairs, offset encoding or
multiple cells per weight. Similarly, weights requiring greater effective precision may be distributed across several devices or bit slices. These encoding choices affect array area, programming cost, peripheral complexity and energy efficiency.
Input values are encoded as voltages, currents, pulse widths or pulse sequences and applied along the rows. The resulting column currents accumulate according to Ohm's and Kirchhoff's laws. In an ideal voltage-input crossbar, the current measured
at column \(j\) is

\begin{equation}
    I_j = \sum_i G_{ij}V_i ,
    \label{eq:ideal-aimc}
\end{equation}

where \(G_{ij}\) is the programmed conductance and \(V_i\) is the encoded input. The array therefore evaluates many multiply--accumulate operations concurrently through its physical electrical response. Following read-out and conversion, a digital nonlinear function may be applied:

\begin{equation}
    Y_j = f\!\left(Q_b(I_j)\right),
    \label{eq:aimc-output}
\end{equation}

where \(Q_b(\cdot)\) represents analogue-to-digital conversion at \(b\)-bit
resolution and \(f(\cdot)\) represents a nonlinear activation or subsequent
digital processing operation.
This spatial parallelism can reduce the repeated movement of weights between physically separate memory and arithmetic units, which is a major source of energy consumption in conventional digital architectures~\cite{horowitz2014computing}.
However, AIMC does not eliminate data movement. Inputs must be delivered to the array; outputs and partial sums must be collected; and large matrices may need to be partitioned across multiple tiles. Routing, buffering and accumulation between tiles can therefore become significant system costs.
\paragraph{Programming and inference}
A distinction is needed between programming the array and using it for inference. During inference, devices are read at relatively low voltages to avoid disturbing their stored states. Programming RRAM or PCM devices may require repeated write pulses and write--verify cycles to bring each conductance sufficiently close to its target value. Programming energy and latency can be substantially higher than those of a read operation. Endurance also limits the number of reliable conductance updates. These considerations make many current non-volatile AIMC systems better suited to inference or infrequent adaptation than to continuous, high-frequency weight updating~\cite{sebastian2020memory,zidan2018future}.

\paragraph{Device and circuit non-idealities}
The ideal relation in Equation~\ref{eq:ideal-aimc} does not fully describe a physical array. Programming error, device-to-device variation, conductance drift,
temperature sensitivity, electrical noise, parasitic wire resistance and converter
quantisation cause the measured output to differ from the ideal result. A simplified
representation is

\begin{equation}
    I^{\mathrm{phys}}_j =
    \sum_i
    \left[
        G_{ij}
        + \Delta G^{\mathrm{prog}}_{ij}
        + \Delta G^{\mathrm{drift}}_{ij}(t,T)
    \right]
    \left(V_i+\eta_i\right)
    + \Delta I^{\mathrm{line}}_j
    + \eta_j ,
    \label{eq:aimc-nonideal}
\end{equation}

followed by

\begin{equation}
    \hat{Y}_j =
    f\!\left(Q_b\!\left(I^{\mathrm{phys}}_j\right)\right).
    \label{eq:aimc-physical-output}
\end{equation}

\begin{figure}[htbp]
\centering
\includegraphics[width=0.90\linewidth]{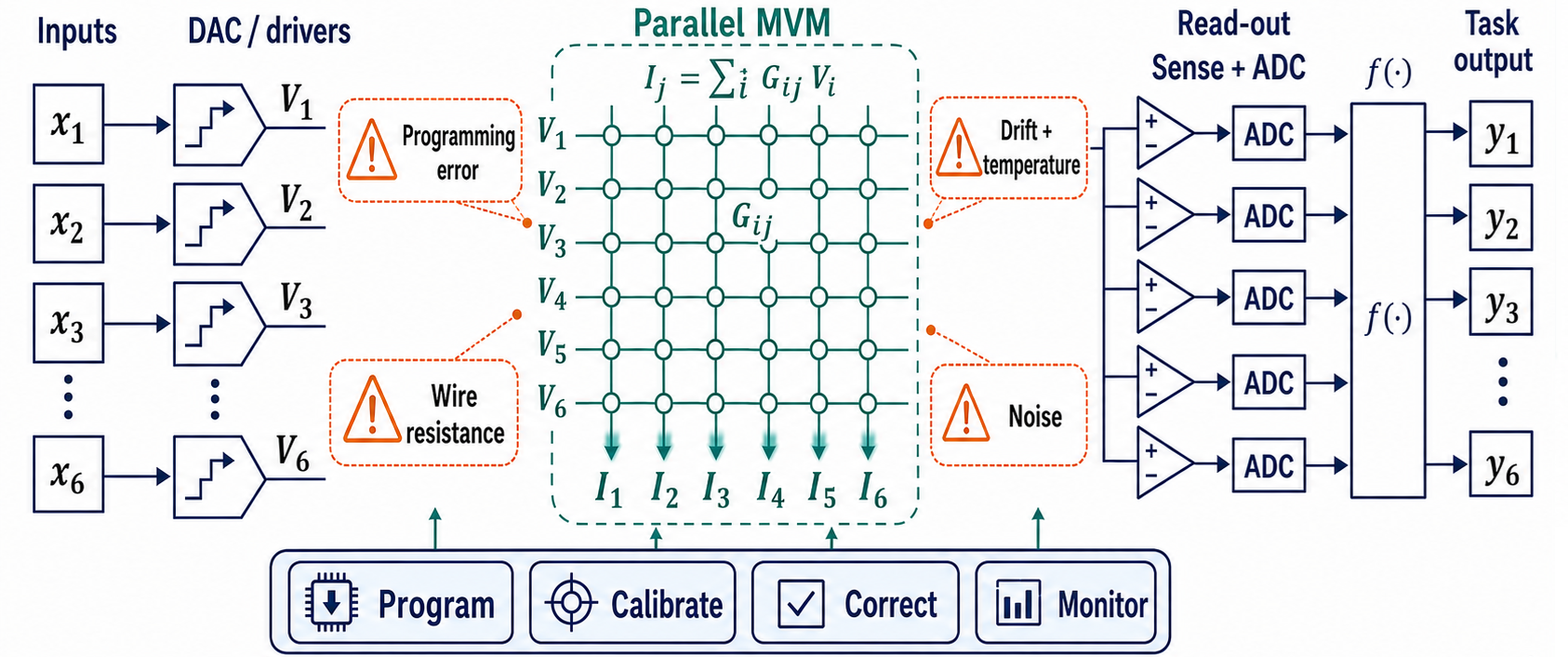}
\caption{Analogue in-memory computing: physical operation and system overheads.}
\label{fig:analogue}
\end{figure}

Here, \(\Delta G^{\mathrm{prog}}_{ij}\) represents the difference between
the target and programmed conductance;
\(\Delta G^{\mathrm{drift}}_{ij}(t,T)\) represents time- and
temperature-dependent conductance variation;
\(\Delta I^{\mathrm{line}}_j\) captures errors associated with wire resistance
and voltage drop; and \(\eta_i\) and \(\eta_j\) represent input- and
output-referred noise. Converter quantisation and saturation are represented
through \(Q_b(\cdot)\). The magnitude and behaviour of these terms depend on the
device technology, array architecture, operating conditions and encoding
scheme~\cite{gallo2018mixed,nandakumar2018phase}.
Array-level accuracy can become more difficult to maintain as arrays grow because
wire resistance, accumulated device variation, noise and limited converter
resolution interact. There is no universal maximum array dimension or
signal-to-noise threshold. A viable tile size depends on the device, circuit,
encoding, workload, utilisation and error budget. Larger arrays provide greater
parallelism and may amortise some peripheral costs, but they can intensify
electrical non-idealities and increase routing complexity. Tile size is therefore a
hardware--algorithm co-design variable rather than a fixed design rule.
\paragraph{Managing physical uncertainty}
AIMC systems can respond to physical uncertainty at several levels. Device-level methods include iterative write--verify programming, conductance-state selection
and redundant device encoding. Circuit-level methods include differential
representation, reference columns, current limiting, adaptive sensing and
calibration. Algorithm-level methods include hardware-aware training, noise
injection, quantisation-aware training, retraining with measured device
characteristics and mixed-precision correction~\cite{gallo2018mixed}.
Digital correction can be used to refine uncertain analogue results. In a
mixed-precision arrangement, the AIMC array performs the high-volume approximate
operation, while a digital processor estimates or corrects the residual error. The
benefit depends on whether the correction cost remains smaller than executing the
complete operation digitally. Excessive correction, repeated reprogramming or
frequent calibration can remove the original AIMC energy advantage.
Conductance drift is particularly relevant to PCM and some resistive-memory
technologies, but its magnitude and operational importance vary across devices.
Digital control blocks may monitor temperature and output statistics, apply
calibration coefficients, inject reference vectors or trigger refresh and
reprogramming. Such operations may occur periodically, when a monitored threshold
is exceeded, or during scheduled maintenance. They should not be assumed to run
continuously in every AIMC system.
This division of responsibility illustrates the role of hybrid computing. The
analogue array provides parallel, data-local computation, while the digital layer
provides configuration, scheduling, nonlinear processing, calibration, correction
and fallback. Hybrid edge classifiers combining digital neural processing with
RRAM-based analogue matching demonstrate this broader principle, although their
specific architecture differs from a conventional MVM crossbar
~\cite{woodward2026hybrid}.
As illustrated in Figure~\ref{fig:analogue}, input values are converted
into voltages applied to the crossbar, where conductance values perform parallel
matrix--vector multiplication. The complete operation also requires read-out,
data conversion, digital processing, calibration and correction.

\paragraph{Conversion and peripheral overhead}
An AIMC tile requires more than the conductance array. It may include input
drivers or digital-to-analogue converters, row and column selection circuits,
sample-and-hold stages, current sensing, analogue-to-digital converters,
multiplexers, buffers, routing and digital control. Depending on precision,
throughput and array utilisation, these peripheral components can account for a
substantial proportion of the tile's energy, area and latency.
At higher output resolution or low array utilisation, conversion can dominate the
tile energy budget. Several approaches can reduce this cost. These approaches introduce trade-offs among precision, throughput, latency,
circuit area and robustness. Lower converter precision, for example, reduces
conversion cost but may increase task error or require additional digital
correction.
A simplified tile-energy decomposition is
\begin{equation}
\begin{split}
    E_{\mathrm{tile}} =\;&
    E_{\mathrm{array}}
    + N_{\mathrm{out}}E_{\mathrm{ADC}}(b_{\mathrm{out}})
    + N_{\mathrm{in}}E_{\mathrm{drive}}(b_{\mathrm{in}}) \\
    &+ E_{\mathrm{buffer}}
    + E_{\mathrm{control}}
    + E_{\mathrm{routing}}
    + E_{\mathrm{calibration}},
\end{split}
\label{eq:aimc-tile-energy}
\end{equation}

where \(N_{\mathrm{in}}\) and \(N_{\mathrm{out}}\) are the numbers of active
input and output channels, while \(b_{\mathrm{in}}\) and \(b_{\mathrm{out}}\)
represent their respective effective precisions. The terms may be shared,
amortised or activity-dependent; Equation~\ref{eq:aimc-tile-energy} is therefore an accounting boundary rather than a universal circuit model.
The equation makes clear that reducing \(E_{\mathrm{array}}\) alone produces
diminishing system-level returns if conversion, routing, buffering or calibration remain dominant. Peripheral-aware design, array utilisation and workload mapping must therefore be treated as first-class objectives in AIMC
hardware--algorithm co-design~\cite{sebastian2020memory,gallo2018mixed}.
The appropriate comparison is between a complete AIMC-based system and a strong digital implementation at matched task accuracy, latency and operating conditions.
A core-level advantage is meaningful only when it remains after input encoding, conversion, inter-tile communication, digital correction, calibration and programming costs have been included. 
\subsubsection{Photonic Processing}
\begin{figure}[htbp]
    \centering
    \includegraphics[width=\linewidth]
    {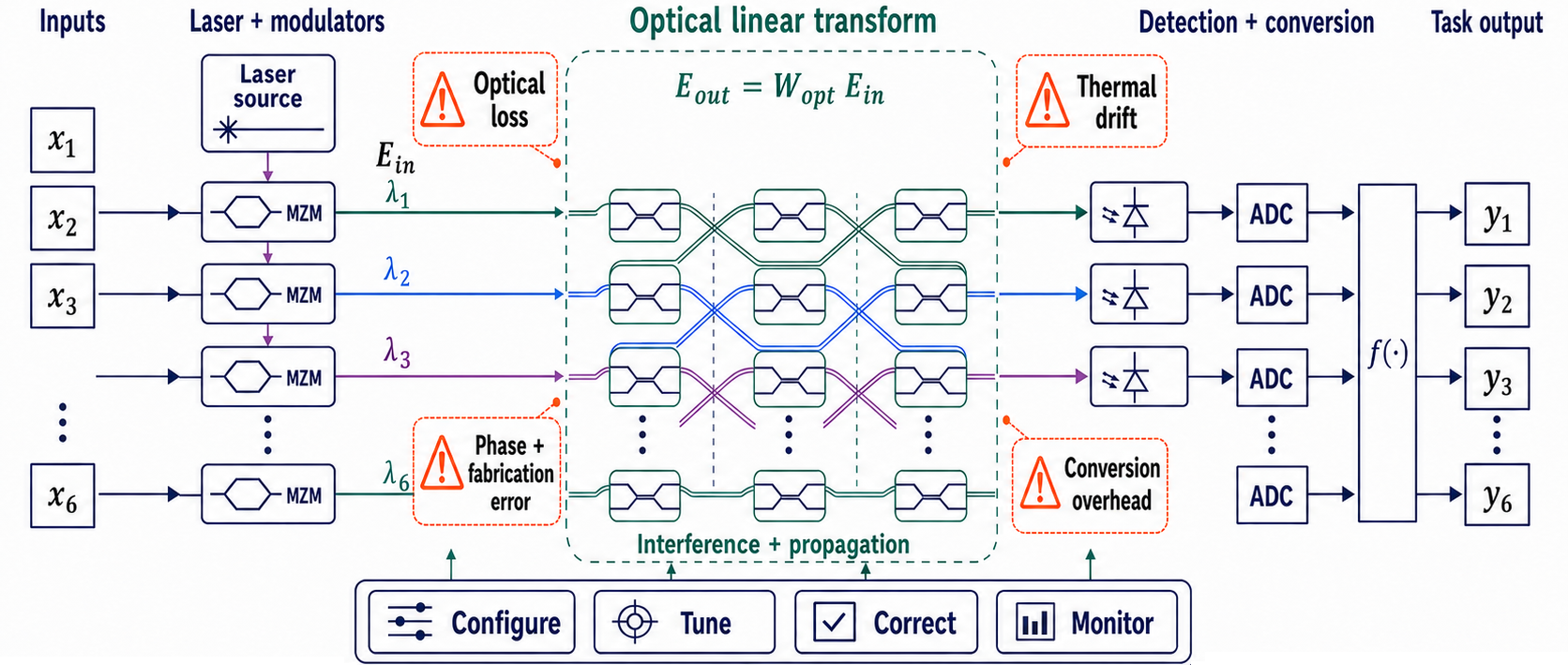}
    \caption{System-level operation of photonic computing. Electronic inputs
    modulate optical carriers that pass through a configurable interferometric
    network to perform a linear transformation. Photodetection, data conversion
    and digital nonlinear processing remain necessary. Optical loss,
    fabrication and phase errors, thermal drift and conversion overhead affect
    the realised system-level performance.}
    \label{fig:photonic-computing}
\end{figure}

Photonic computing uses the propagation, interference and modulation of light to perform selected mathematical operations. Its principal attraction lies in the high bandwidth of optical signals and the ability to encode and process information across multiple physical dimensions, including amplitude, phase, wavelength, time, polarisation and spatial mode~\cite{fu2024optical}. These properties make photonic hardware particularly relevant to linear algebra, convolution, Fourier transforms, signal processing and other workloads with substantial parallel structure.
Photonic processors do not replace the complete digital system. Electronic data must generally be encoded onto optical carriers using modulators, while photodetectors and electronic read-out circuits recover the outputs. Digital processors remain responsible for memory access, workload mapping, control, nonlinear operations, calibration and communication with the wider system. Most practical photonic AI accelerators should therefore be understood as optoelectronic hybrid systems~\cite{bogaerts2020programmable, Cong:26}.
As illustrated in Figure~\ref{fig:photonic-computing}, electronic inputs
modulate optical carriers that pass through a configurable interferometric
network. The optical core performs a linear transformation through interference
and propagation, while light generation, detection, conversion, digital
nonlinear processing and system control remain within the complete system
boundary.

\paragraph{Optical linear transformations}
A photonic processor can represent an input vector using the amplitude and phase of
an optical field. If the encoded input field is denoted by
\(\mathbf{E}_{\mathrm{in}}\), a configured linear optical network implements

\begin{equation}
    \mathbf{E}_{\mathrm{out}}
    =
    \mathbf{W}_{\mathrm{opt}}\mathbf{E}_{\mathrm{in}},
    \label{eq:photonic-linear}
\end{equation}

where \(\mathbf{W}_{\mathrm{opt}}\) is the transformation represented by the
physical optical network. Depending on the architecture, the transformation may be
implemented using Mach--Zehnder interferometers (MZIs), microring resonators,
diffractive elements, wavelength-selective filters or phase-change photonic
devices~\cite{feldmann2021parallel}.

Photodetectors normally measure optical intensity rather than the signed complex
field directly:

\begin{equation}
    z_j \propto
    \left|E_{\mathrm{out},j}\right|^2 .
    \label{eq:photonic-detection}
\end{equation}

Coherent detection, balanced photodetectors or differential encoding may be used
when phase information or signed outputs must be recovered. Following detection,
quantisation and digital processing, the task output may be expressed as

\begin{equation}
    \hat{\mathbf{Y}}
    =
    f\!\left(
        Q_b\!\left[
            \mathcal{D}
            \left(
                \mathbf{W}_{\mathrm{opt}}
                \mathbf{E}_{\mathrm{in}}
            \right)
        \right]
    \right),
    \label{eq:photonic-output}
\end{equation}

where \(\mathcal{D}(\cdot)\) represents optical detection, \(Q_b(\cdot)\)
represents electronic quantisation at \(b\)-bit resolution and \(f(\cdot)\)
represents subsequent nonlinear or digital processing.

\paragraph{Interferometer meshes}

An MZI combines and separates optical signals using controllable phase shifts.
Networks of MZIs can be configured to perform matrix transformations. For an
\(N\times N\) unitary transformation, the rectangular Clements architecture
requires tunable \(2\times2\) interferometric elements~\cite{clements2016optimal}. 

\begin{equation}
    M_{\mathrm{MZI}}
    =
    \frac{N(N-1)}{2}
    \label{eq:mzi-count}
\end{equation}

The rectangular arrangement offers balanced optical paths and lower optical depth than some earlier triangular decompositions.
Equation~\ref{eq:mzi-count} applies to a unitary transformation; it does not mean
that any arbitrary matrix can be implemented using a single mesh of this size. A
general complex matrix \(\mathbf{W}\) may be expressed through singular-value
decomposition:

\begin{equation}
    \mathbf{W} = \mathbf{U} \bm{\Sigma} \mathbf{V}^{\dagger}
    \label{eq:photonic-svd}
\end{equation}

where \(\mathbf{U}\) and \(\mathbf{V}^{\dagger}\) are unitary transformations and $\bm{\Sigma}$ contains the singular values. A corresponding physical implementation may require two interferometer meshes and an intermediate set of
optical attenuators or amplifiers. Rectangular matrices may require padding, partitioning or alternative architectures. The area, insertion loss, tuning effort and control complexity therefore grow with matrix dimension.

\paragraph{Sources of optical parallelism}

Photonic systems offer several forms of parallelism:

\begin{itemize}
    \item \textbf{Spatial parallelism:} multiple waveguides, free-space paths or
    optical modes process different signal components concurrently.

    \item \textbf{Wavelength-division multiplexing:} multiple wavelength channels
    can carry and process independent values through a shared optical path.

    \item \textbf{Temporal multiplexing:} high-rate optical modulation allows
    sequences of values to pass through the same physical operator.

    \item \textbf{Propagation-based computation:} a configured network performs
    its linear transformation as the optical signal propagates through it.
\end{itemize}

The number of usable wavelength, spatial and temporal channels is finite. It is
limited by optical bandwidth, channel spacing, crosstalk, detector bandwidth,
modulation rate, signal-to-noise ratio and available optical power. Wavelength
multiplexing should therefore be described as high parallelism rather than an
unlimited number of computational lanes.

The optical transformation itself can have low latency because it occurs during
propagation. However, end-to-end latency includes memory access, modulation,
optical propagation, detection, conversion, buffering and digital control.
Furthermore, optical path length, component count and routing complexity grow with
the physical architecture. Photonic computation should not therefore be described
as having universally constant \(\mathcal{O}(1)\) task latency.

\paragraph{Energy and interface costs}

Passive optical interference does not require transistor switching within the
transformation itself. Nevertheless, a complete photonic processor requires energy
for light generation, input modulation, weight programming or tuning, detection,
electronic conversion and control. A system-level task-energy boundary can be
written as

\begin{equation}
\begin{split}
    E_{\mathrm{photonic,task}} =\;&
    E_{\mathrm{memory}}
    + E_{\mathrm{laser}}
    + E_{\mathrm{modulation}}
    + E_{\mathrm{optical,core}} \\
    &+ E_{\mathrm{tuning}}
    + E_{\mathrm{detection}}
    + E_{\mathrm{conversion}}
    + E_{\mathrm{digital}} .
\end{split}
\label{eq:photonic-energy}
\end{equation}

For a mostly passive optical core, \(E_{\mathrm{optical,core}}\) may be relatively small, but this does not imply zero-energy computation. Laser wall-plug efficiency, optical loss, modulator drive energy, detector sensitivity and ADC/DAC requirements can dominate the complete energy budget. The optical core must also maintain sufficient signal power as the network depth, fan-out and number of channels increase.

Photonics is consequently most attractive when the optical operator is used at high throughput and utilisation, allowing light-source, modulation, tuning and conversion costs to be amortised across many useful operations~\cite{hua2025integrated}. Repeated transfer between electronic and optical domains can remove the advantage of the optical core, particularly for small workloads or frequently interrupted execution~\cite{10.1093/nsr/nwag089}.

\paragraph{Precision, noise and optical loss}
Photonic computing is analogue and therefore subject to noise and physical
variation, including laser noise, shot noise, thermal noise,
photodetector noise, phase error, crosstalk and fabrication mismatch. The optical power budget becomes increasingly important as signals pass through multiple components or are divided across several outputs. Greater numerical precision generally requires improved signal-to-noise ratio, greater optical power, more accurate phase control or additional electronic correction.

\paragraph{Thermal drift and calibration}
MZIs and microring resonators are sensitive to fabrication variation and temperature. Changes in refractive index alter optical phase and resonance wavelength, causing the realised matrix to deviate from its programmed state. Microring architectures can be particularly sensitive because their operation depends on maintaining wavelength alignment between resonators and optical channels ~\cite{biasi2023photonic}.

Thermal heaters, carrier injection, reference signals and feedback control may be used to configure and stabilise the optical circuit. These mechanisms add static or dynamic power, occupy chip area and require electronic monitoring. Non-volatile photonic materials can reduce continuous tuning requirements, but they introduce their own programming, endurance, loss and variability considerations.

\paragraph{Nonlinear operations}
Optical interference and propagation naturally implement linear transformations. Optical nonlinearities exist, but practical implementations may require high optical power, introduce loss, provide limited reconfigurability or be difficult to cascade at scale~\cite{fu2024optical}.

Many current systems therefore execute matrix operations optically and return to electronics for nonlinear activation and control. This arrangement improves
programmability but introduces repeated optical-to-electrical and electrical-to-optical conversion. The number and placement of these boundaries
become important architectural choices. A system that converts after every small operation may lose the bandwidth and energy advantages of the photonic core.

\section{Neuromorphic Computing}

Neuromorphic computing draws inspiration from biological nervous systems to process information using distributed state, temporal dynamics, local memory access and sparse event communication~\cite{indiveri2011neuromorphic,kudithipudi2025scale}.   It is not synonymous with analogue
computing or with spiking neural networks (SNNs). Neuromorphic systems may be fully digital, mixed-signal or based on emerging physical devices, and some support
non-spiking event-driven models alongside biologically inspired neural dynamics.

The principal architectural opportunity is to make computation conditional on activity. Instead of repeatedly evaluating every element of a dense tensor at a fixed rate, an event-driven system can update and communicate only the states affected by an input event \cite{Gabayre25}. This approach is particularly relevant to temporally sparse sensor streams and continuously operating edge systems. However, the benefit depends on the complete implementation: clocking or bias generation, state storage, event routing and input--output processing may consume energy even when few spikes are generated.

\subsection{Neural Dynamics and Event Representation}

A commonly used abstraction is the leaky integrate-and-fire (LIF) neuron. For neuron \(i\), its membrane state may be represented as

\begin{equation}
C_{\mathrm{m}}\frac{\mathrm{d}V_i(t)}{\mathrm{d}t}
=
-g_{\mathrm{L}}\left[V_i(t)-E_{\mathrm{L}}\right]
+ I_i(t),
\label{eq:lif-neuron}
\end{equation}

where $C_{\mathrm{m}}$ is the membrane capacitance,
$g_{\mathrm{L}}$ is the leak conductance,
$E_{\mathrm{L}}$ is the resting potential and
$I_i(t)$ is the input current.

\begin{equation}
\text{if } V_i(t)\geq V_{\mathrm{th}},
\qquad
S_i(t)=1,
\qquad
V_i(t)\leftarrow V_{\mathrm{reset}} .
\label{eq:lif-threshold}
\end{equation}

A presynaptic spike train can be written as

\begin{equation}
S_j(t)=\sum_k \delta\!\left(t-t_j^{(k)}\right),
\label{eq:neuromorphic-spike-train}
\end{equation}
where $t_j^{(k)}$ is the time of the $k$-th spike emitted by neuron $j$.

\begin{equation}
I_i(t)
=
\sum_j w_{ij}\left(h\ast S_j\right)(t),
\label{eq:neuromorphic-synaptic-current}
\end{equation}

where $w_{ij}$ is the synaptic weight, $h(t)$ is a temporal
response kernel and $\ast$ denotes convolution.

The LIF model is only one possible abstraction. Adaptive exponential models
support adaptation and bursting~\cite{brette2005adex}, while resonant models
capture subthreshold oscillations and frequency-selective responses
~\cite{izhikevich2001resonate}. Conductance-based and multi-compartment models
provide richer representations of synaptic and dendritic dynamics
~\cite{gerstner2014neuronal}. These more expressive models may improve temporal
processing but increase state-storage, arithmetic and calibration requirements.
The appropriate neuron model is therefore an algorithm--hardware co-design
choice; greater biological detail is not inherently preferable.

\begin{figure*}[t]
    \centering
    \includegraphics[width=\textwidth]
    {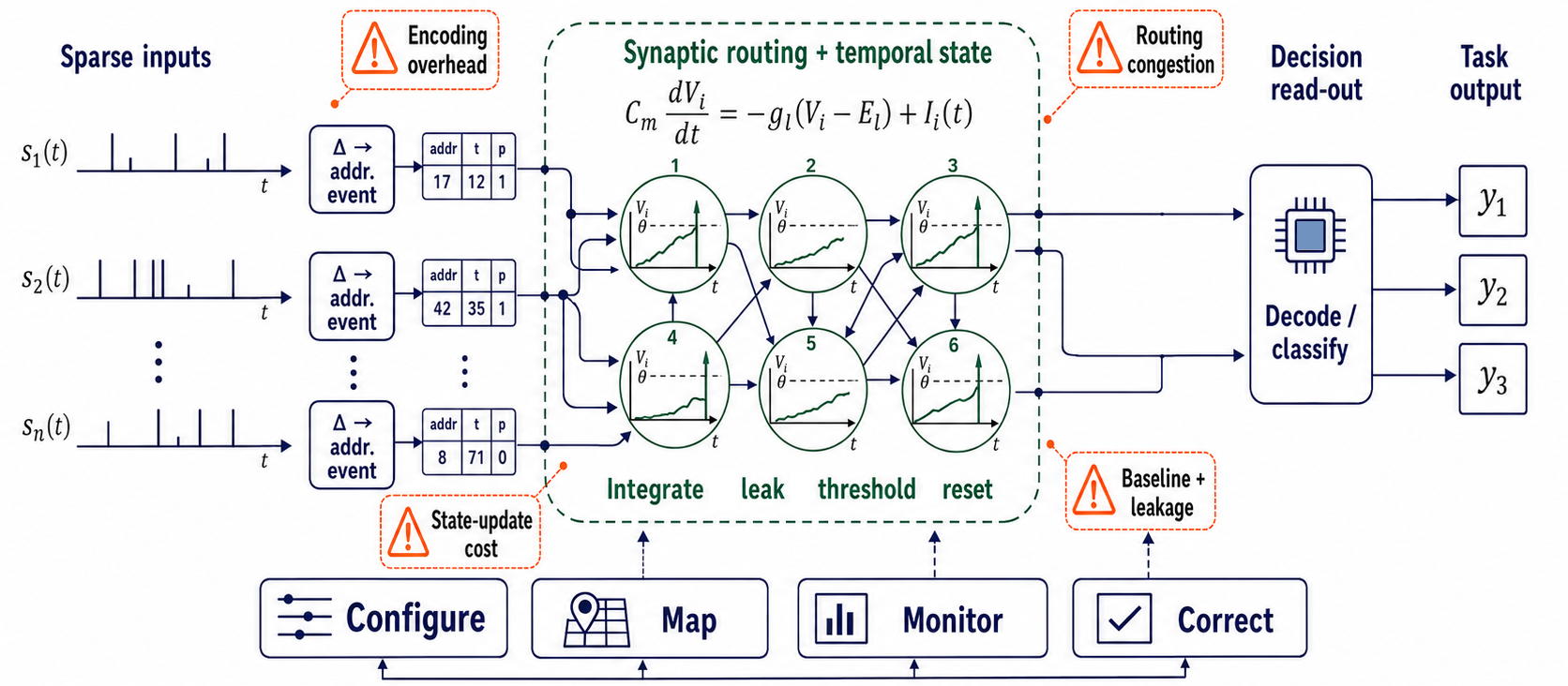}
    \caption{System-level view of neuromorphic computing. Sparse input events
    are encoded and routed through a network of stateful neurons, where membrane
    states integrate, leak, cross a threshold and reset. Sparse activity can
    reduce computation and communication, but encoding, state updates, routing,
    memory, baseline power and system control remain part of the task-level
    energy cost.}
    \label{fig:neuromorphic}
\end{figure*}

Figure~\ref{fig:neuromorphic} illustrates the event-driven processing
path and the wider costs that must be included when evaluating a neuromorphic
system~\cite{indiveri2011neuromorphic,kudithipudi2025scale,
yik2025neurobench}.

\subsection{Activity-Dependent Energy}

Sparse activity can reduce switching, synaptic processing and communication, but spike count alone does not determine system energy. A more useful task-level decomposition is

\begin{equation}
\begin{split}
E_{\mathrm{task}} ={}&
P_{\mathrm{base}}T
+ N_{\mathrm{syn}}E_{\mathrm{syn}}
+ N_{\mathrm{spk}}E_{\mathrm{route}} \\
&+ N_{\mathrm{upd}}E_{\mathrm{state}}
+ E_{\mathrm{memory}}
+ E_{\mathrm{I/O}}
+ E_{\mathrm{host}} .
\end{split}
\label{eq:neuromorphic-task-energy}
\end{equation}

where $P_{\mathrm{base}}$ is the baseline power over execution time $T$,
$N_{\mathrm{syn}}$ is the number of delivered synaptic events,
$N_{\mathrm{spk}}$ is the number of routed spike packets, and
$N_{\mathrm{upd}}$ is the number of neuron-state updates. The remaining
terms account for memory access, sensor and interface processing, and
host-side control or preprocessing.

This expression is an accounting boundary rather than a universal device model. In a clock-driven digital implementation, state-update energy may depend on the simulation time step even when no spikes occur. In an asynchronous digital implementation, activity may more directly determine switching and routing. In a mixed-signal implementation, physical circuits may evolve neuron states continuously, but bias currents, leakage, calibration and communication still contribute to total energy.

Neuromorphic execution is therefore most advantageous when useful information can be represented with sufficiently few state updates and communicated events, and when the reduction in activity exceeds the associated baseline and interface costs. Energy per synaptic operation is informative for characterising a core, but energy per completed task is required for comparison with conventional digital systems~\cite{ostrau2022benchmark,yik2025neurobench}.

\subsection{Training and Mapping}

SNNs are a type of neural networks that can be defined on LIF neuron structures defined above \cite{Illeperuma24}. They are specifically designed to process spike trains. To run ANNs on spiking neural processors, the equivalent SNNs may be obtained through conversion from trained artificial neural networks, trained directly using surrogate gradients, or developed using local and event-driven learning rules~\cite{neftci2019surrogate}. These approaches have different hardware consequences. ANN-to-SNN conversion can preserve familiar training workflows but may require many time steps or high firing rates to reproduce the original activation values. Direct temporal training can exploit spike timing and recurrent state but introduces optimisation and deployment complexity.

Hardware-aware training can incorporate limits on weight precision, fan-in, neuron parameters, routing capacity and permissible time constants. These constraints affect task accuracy, latency and event activity simultaneously. A model that is accurate in software may lose its energy advantage after mapping if it requires excessive spike rates, repeated time steps, off-chip state access or unsupported operations.

Software interoperability is also an open challenge because neuromorphic platforms expose different neuron models, timing semantics and connectivity constraints. The Neuromorphic Intermediate Representation provides one approach for expressing continuous-time neuromorphic computations across heterogeneous platforms~\cite{pedersen2024nir}.

\subsection{Hardware Diversity}

Digital neuromorphic processors represent neuron states, synaptic weights and events using digital logic and memory. Examples include SpiNNaker, SpiNNaker2 and Intel Loihi~2~\cite{furber2014spinnaker,gonzalez2024spinnaker,orchard2021loihi2}. Digital implementations generally provide reproducible numerical behaviour and flexible programmability, although memory access, discrete state updates and packet routing remain significant energy costs.

Mixed-signal systems use analogue circuits for some neural or synaptic dynamics while retaining digital communication, configuration and control. BrainScaleS-2, for example, implements accelerated analogue neuron and synapse dynamics together with digital event handling and programmable processors~\cite{pehle2022brainscales}. DYNAP-SE2 similarly combines analogue neural dynamics with asynchronous digital event routing~\cite{richter2024dynapse2}. These systems can exploit physical dynamics directly, but device mismatch, operating-point variation and calibration become part of the programming model.

\subsection{Workload Suitability and Evaluation}
Neuromorphic architectures are strongest candidates for workloads with persistent temporal structure and sparse changes, including event-based vision, always-on audio, radar, tactile sensing, biosignal analysis, streaming anomaly detection and closed-loop control. Event cameras are relevant input devices because they report local intensity changes asynchronously rather than transmitting complete frames~\cite{gallego2022event}. They are sensors rather than neuromorphic processors, and the energy required for sensing, event encoding and data transfer must remain inside the evaluation boundary.
Benefits are less certain when dense static inputs must first be converted into long spike sequences, when high firing rates are required to maintain accuracy, or when unsupported operations repeatedly return execution to a host processor. Latency must also be stated carefully: a fast individual event response does not necessarily imply low task latency if a decision requires a long observation window.
Neuromorphic systems should therefore be evaluated using (i) task accuracy, (ii) energy per task, (iii) end-to-end latency, (iv) event activity, (v) memory traffic, (vi) sensor and host overhead and (vii) robustness under deployment conditions. Benchmarking frameworks such as NeuroBench are valuable because they separate algorithm-level characteristics from measurements made on physical systems and encourage comparisons under declared workload and measurement boundaries~\cite{yik2025neurobench}.

\section{System-Level Challenges and Non-Idealities}

The advantages of analogue in-memory, photonic and neuromorphic computing arise from specialised physical operations. This dependence on physical behaviour also introduces variability, environmental sensitivity and interface costs. The output of a physical accelerator can be represented conceptually as

\begin{equation}
\hat{\mathbf{y}}
\approx
\mathbf{y}_{\mathrm{ideal}}
+ \boldsymbol{\varepsilon}_{\mathrm{device}}
+ \boldsymbol{\varepsilon}_{\mathrm{circuit}}
+ \boldsymbol{\varepsilon}_{\mathrm{conversion}}
+ \boldsymbol{\varepsilon}_{\mathrm{mapping}}
+ \boldsymbol{\varepsilon}_{\mathrm{environment}} .
\label{eq:system-error-decomposition}
\end{equation}

where the error terms represent device variation, circuit non-idealities, conversion and quantisation, hardware mapping and environmental effects such as temperature, supply variation and ageing. These terms may interact and are not necessarily independent or additive; the expression provides an accounting framework rather than a universal error model.

In AIMC, programming variation, conductance drift, read noise, limited conductance range and line resistance can alter the effective weights~\cite{gallo2018mixed,nandakumar2018phase,sebastian2020memory}. Photonic systems are affected by optical loss, detector noise, fabrication variation, wavelength instability and thermal phase drift~\cite{biasi2023photonic,shekhar2024silicon}. Neuromorphic systems may experience parameter mismatch, timing variation, limited fan-in, state-update costs and routing congestion. There is no universal effective precision for these platforms: achievable accuracy depends on the device, circuit, encoding, array size, operating conditions and workload tolerance.

The physical core is only one part of the system. Inputs must be retrieved or sensed, encoded and transferred to the accelerator, while outputs must be detected, converted, buffered and communicated. These costs can dominate when the core is small, poorly utilised or interrupted by unsupported operations. Moving weights closer to computation reduces weight movement but does not remove the transfer of inputs, activations, partial sums, events and control information. Optical systems still require lasers, modulation, detection and electronic memory, while neuromorphic systems retain baseline power, state maintenance, routing and event-encoding costs.

Programming and calibration introduce further overhead. AIMC devices may require iterative programme-and-verify operations; photonic circuits may require phase tuning, wavelength alignment and thermal stabilisation; and neuromorphic processors require models to be mapped across finite neuron, synapse, memory and routing resources. Calibration consumes energy and time, and its cost depends on how frequently it is performed and how long the calibrated state remains valid. Continual adaptation may also be constrained by programming energy, device endurance, update asymmetry and limited on-chip learning support.

Scaling introduces device, circuit and packaging level challenges. Larger AIMC arrays intensify line resistance, accumulated variation and peripheral loading. Photonic systems face optical loss, phase stability, laser integration and electrical--optical packaging constraints~\cite{shekhar2024silicon}. Neuromorphic systems need to manage event-routing capacity, distributed memory and communication between chips. Manufacturing yield, device uniformity and integration with established CMOS processes remain important, although the severity of these constraints depends on the technology and foundry process.

Software must expose hardware constraints such as precision, array dimensions, supported neural dynamics, routing capacity and calibration state. A model that performs well in simulation may behave differently after physical mapping. Intermediate representations such as the Neuromorphic Intermediate Representation support portability across several neuromorphic platforms, but broader interoperability remains unresolved~\cite{pedersen2024nir}. Reproducible evaluation should therefore report the hardware revision, mapping method, operating conditions, calibration procedure, host involvement and measurement boundary~\cite{yik2025neurobench}.

\section{Towards Hybrid Digital--Analogue Computing}

Energy-efficient AI does not require the complete replacement of digital
infrastructure. A hybrid system assigns suitable operations to specialised
physical substrates while retaining digital computation for memory management,
runtime scheduling, nonlinear operations, communication, security and result
verification. Effective deployment therefore requires co-design across devices,
circuits, numerical representations, algorithms, compilers and runtime
control~\cite{cai2020once,woodward2026hybrid}.

\subsection{Mixed-Signal Integration and Interface Costs}

Mixed-signal circuitry forms the operational boundary between physical signal
representation and digital computation. It is not a separate computing paradigm.
Information may enter a physical core from a digital system or directly from the
environment through a sensor or optical front end. A physical core may represent
information through conductance, charge, voltage, current, optical amplitude,
phase or event timing. Digital circuitry may configure the core, supply encoded
data, reconstruct its outputs and connect it to the wider system. In sensor-native
systems, however, physical processing may occur before digitisation.

A general mixed-signal processing path can be written as

\begin{equation}
    \mathbf{x}_{\alpha}
    \xrightarrow{\mathcal{I}_{\alpha\rightarrow p}}
    \mathbf{s}_{p}
    \xrightarrow{\mathcal{P}_{\theta}}
    \mathbf{r}_{p}
    \xrightarrow{\mathcal{O}_{p\rightarrow\beta}}
    \hat{\mathbf{y}}_{\beta},
    \qquad
    \alpha,\beta\in\{\mathrm{digital},\mathrm{physical}\},
    \label{eq:mixed-signal-path}
\end{equation}

where $\mathcal{I}_{\alpha\rightarrow p}$ denotes the input interface, $\mathcal{P}_{\theta}$ denotes the physical operation configured by parameters $\theta$, and $\mathcal{O}_{p\rightarrow\beta}$ denotes the output interface.
For a digitally driven accelerator, the input interface may include encoding, DACs, drivers or optical modulators. For sensor-native processing, the input may already be physical, and the interface may consist only of transduction, conditioning or direct optical propagation. In computational imaging, for example, an incident optical field may undergo a physical transformation before photodetection and digitisation. The output interface may include sensing, amplification, detection, ADCs or continued processing in another physical domain.

The interface may contain DACs, voltage or current drivers,
optical modulators, sense or transimpedance amplifiers, sample-and-hold circuits, comparators, ADCs, multiplexers and signal-conditioning circuits. These components determine the usable precision, bandwidth, latency and energy of the
accelerator~\cite{aguirre2024hardware}.

\begin{figure*}[t]
    \centering
    \includegraphics[width=\textwidth]{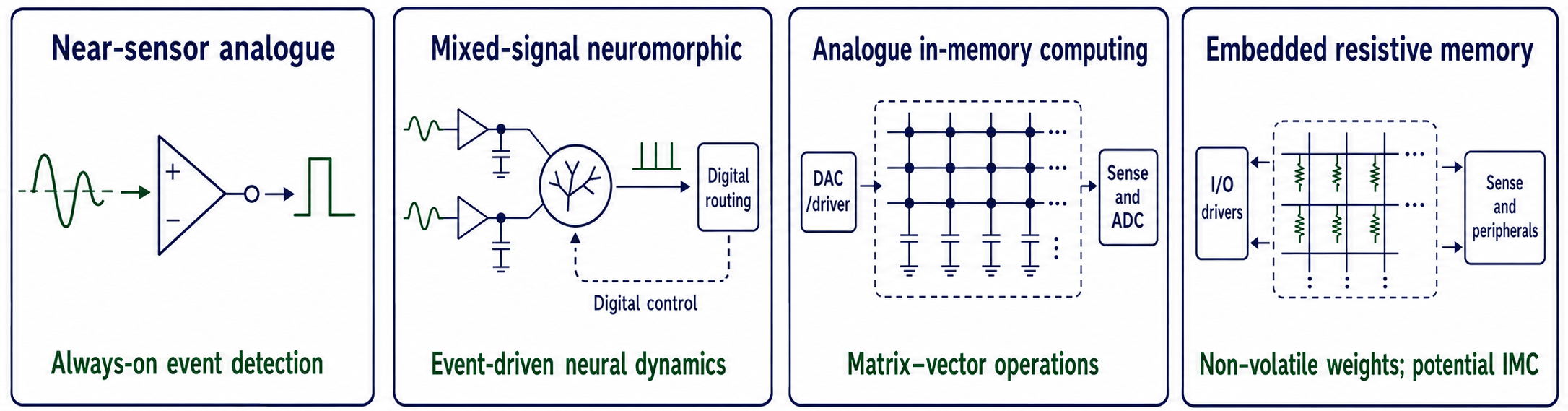}
    \caption{Representative technologies and circuit roles associated with
    physical--digital integration. Inputs may be supplied by a digital system or
    arrive directly as sensor, electrical or optical signals. Physical processing
    may occur after digital encoding or before digitisation. The examples represent
    separate implementations or enabling components rather than parts of a single
    system.}
    \label{fig:mixed-signal-implementations}
\end{figure*}

Figure~\ref{fig:mixed-signal-implementations} distinguishes several forms of
mixed-signal implementation. IBM's PCM-based AIMC chips combine analogue
matrix--vector multiplication with digital activation, communication and
control~\cite{legallo2023mixed,ambrogio2023analog}. Capacitor-based AIMC developed
by EnCharge AI and embedded-flash matrix processing developed by Mythic use
different devices but retain the need for input control, read-out and digital
integration~\cite{lee2021capacitor,mythic2021amp}. Innatera's Pulsar combines an
analogue/mixed-signal spiking engine with digital processing, while SynSense's
DYNAP-SE2 combines analogue neuron and synapse dynamics with asynchronous digital
event routing~\cite{innatera2025pulsar,richter2024dynapse2}. Aspinity places
analogue processing near continuously monitored sensors to determine when a
higher-power digital pipeline should be activated~\cite{aspinity2024analogml}.
Weebit Nano's embedded ReRAM occupies a different level of the stack: it is an
enabling memory technology rather than a complete processor, and still requires
peripheral circuits, calibration, communication and software support for use in
a computing system~\cite{weebit2024reram}.

Converter resolution creates a central trade-off. Greater resolution can reduce
quantisation error but usually increases area, energy or latency. Converter
sharing reduces area but introduces serialisation and may limit throughput.
Digital control may also configure weights or phase settings, monitor drift and
temperature, apply correction factors and schedule recalibration. Hardware-aware
training can improve robustness to limited precision and physical variation, but
correction, repeated measurement and retraining consume resources
~\cite{rasch2023hardware}.

The complete energy boundary is therefore:
\begin{equation}
    E_{\mathrm{mixed}} =
    E_{\mathrm{source}}
    + E_{\mathrm{input}}
    + E_{\mathrm{physical}}
    + E_{\mathrm{output}}
    + E_{\mathrm{digital}}
    + E_{\mathrm{control}}
    + E_{\mathrm{communication}},
    \label{eq:mixed-signal-energy}
\end{equation}

Here, $E_{\mathrm{source}}$ accounts for sensing or signal acquisition when it
lies within the measurement boundary. The input-interface term may include
digital encoding, conversion, modulation, transduction or signal conditioning,
while the output-interface term may include detection, amplification, sensing
and conversion. Individual terms may be absent when the signal enters or leaves
the physical core without crossing a digital--physical boundary.

This boundary explains why results reported for an array, a compute macro, a
processor and a complete application are not directly comparable. The number
and placement of physical--digital boundaries are architectural choices.
Repeated conversion around small operations can remove the advantage of the
physical core, whereas retaining compatible operations in the physical domain
can amortise encoding, conversion and communication costs. Mixed-signal integration concerns the interfaces and coupled physical--digital operation within a component or processing stage. Hybrid design acts at a higher architectural level by selecting and coordinating distinct computing engines, which may themselves contain mixed-signal circuitry.

\subsection{Hybrid System Architecture}
Hybrid digital--analogue computing combines digital and physical processing
within a coordinated system, assigning each the operations for which it is
suited. These engines may use analogue electronic, photonic, in-memory or neuromorphic principles. They may be integrated on the same chip, connected as chiplets or implemented as separate components. The defining feature is coordinated computation across different representations or processing mechanisms, rather than their physical separation.

Mixed-signal and hybrid computing therefore describe different levels of a
system. Mixed-signal refers to "the circuit-level coexistence and conversion of
physical and digital signal representations". Hybrid computing describes "how
computational responsibilities are divided between digital and physical engines".
An analogue accelerator is one form of hybrid computing, in which a digital host delegates selected operations to a physical core. Near-sensor, in-sensor and optical front-end processing provide another form, in which physical computation occurs before digitisation. A system may consequently contain mixed-signal circuitry without constituting a hybrid computing architecture, whereas most hybrid digital--analogue systems require mixed-signal interfaces.

\begin{figure*}[t]
    \centering
    \includegraphics[width=\textwidth]{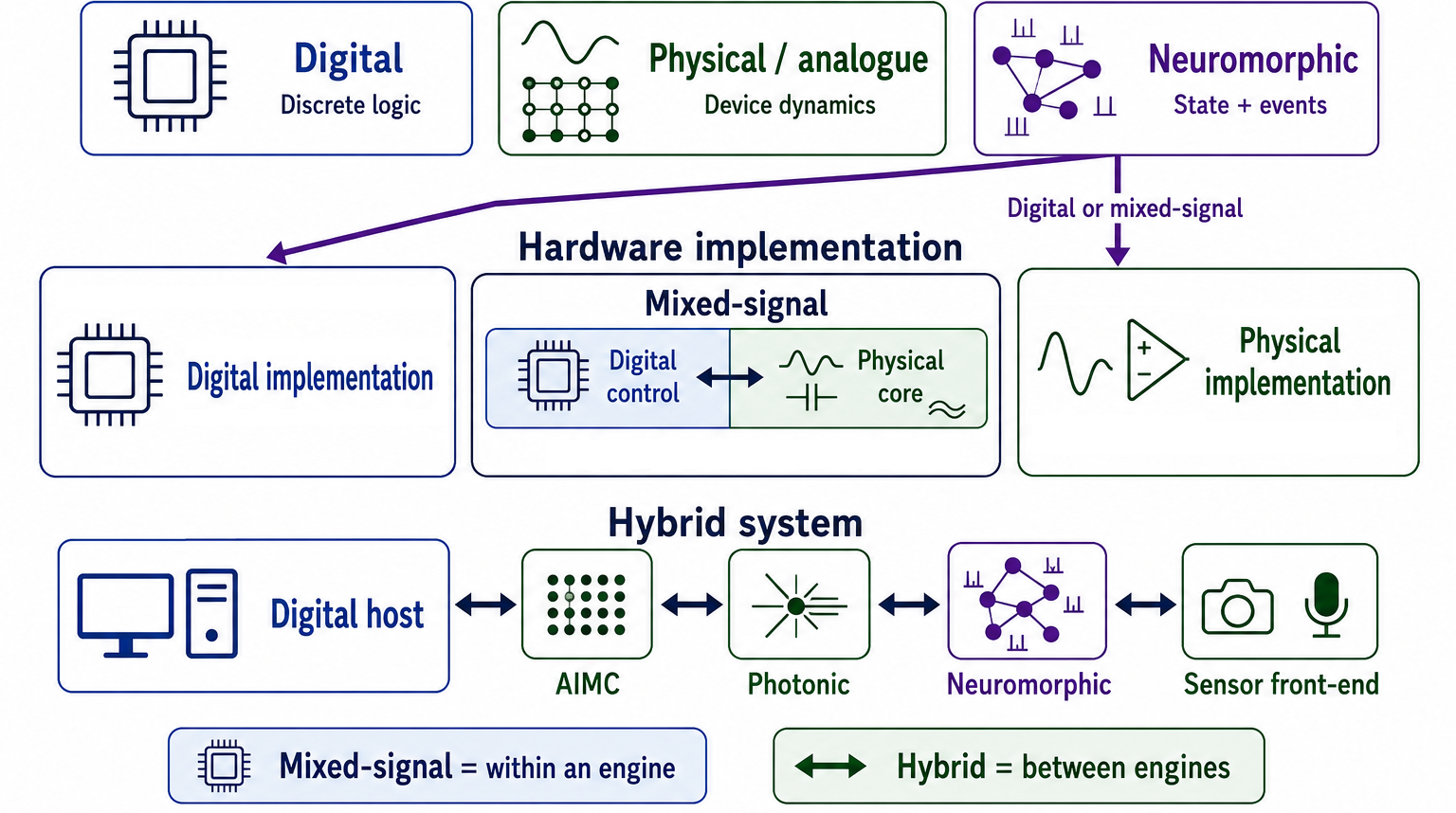}
    \caption{Relationship between digital, physical and neuromorphic computing.
    Neuromorphic systems may use digital, analogue or mixed-signal hardware.
    Mixed-signal describes the integration of physical signal processing with
    digital conversion, control or correction. Hybrid computing describes the
    higher-level coordination of digital and physical computational roles,
    including analogue acceleration and sensor-first physical processing.}
    \label{fig:hybrid0}
\end{figure*}

The technologies also overlap without being interchangeable. A neuromorphic
processor may be implemented using digital, analogue or mixed-signal hardware,
while AIMC and photonic accelerators may execute conventional non-spiking
neural-network operations. Figure~\ref{fig:hybrid0} illustrates these
relationships.

A deployable hybrid architecture can be described through five functional
layers:

\begin{itemize}
    \item \textbf{Sensing and input acquisition:} obtains data from conventional
    sensors, event-based sensors, acoustic arrays, imaging systems or multimodal sensing platforms. The acquired signal may be digitised immediately or retained in a physical form for early processing.

    \item \textbf{Physical processing and specialised acceleration:} performs
    suitable matrix, signal or temporal operations using AIMC, photonic,
    analogue or neuromorphic engines. Physical processing may occur before
    digitisation, as in near-sensor or optical front ends, or after a digital
    processor delegates a suitable computational kernel.

    \item \textbf{Mixed-signal interfaces and data movement:} manages
    transduction, encoding, modulation, sensing, conversion, buffering, routing
    and communication between physical and digital domains.

    \item \textbf{Digital orchestration:} schedules workloads, manages memory,
    configures physical engines, executes unsupported operations and coordinates
    models, devices and communication.

    \item \textbf{Monitoring and assurance:} observes operating conditions,
    detects faults or excessive uncertainty, applies correction and provides a
    digital fallback when required.
\end{itemize}

These layers define functional responsibilities rather than a fixed processing
order. Physical processing may precede digitisation, operate as an accelerator
within a digital pipeline or appear at several points in the system. Multiple
functions may be integrated into one system-on-chip, distributed across chiplets
or separated between an edge device and a host processor.

\subsection{Hybrid Design Patterns}

\begin{figure*}[t]
    \centering
    \includegraphics[width=\textwidth]{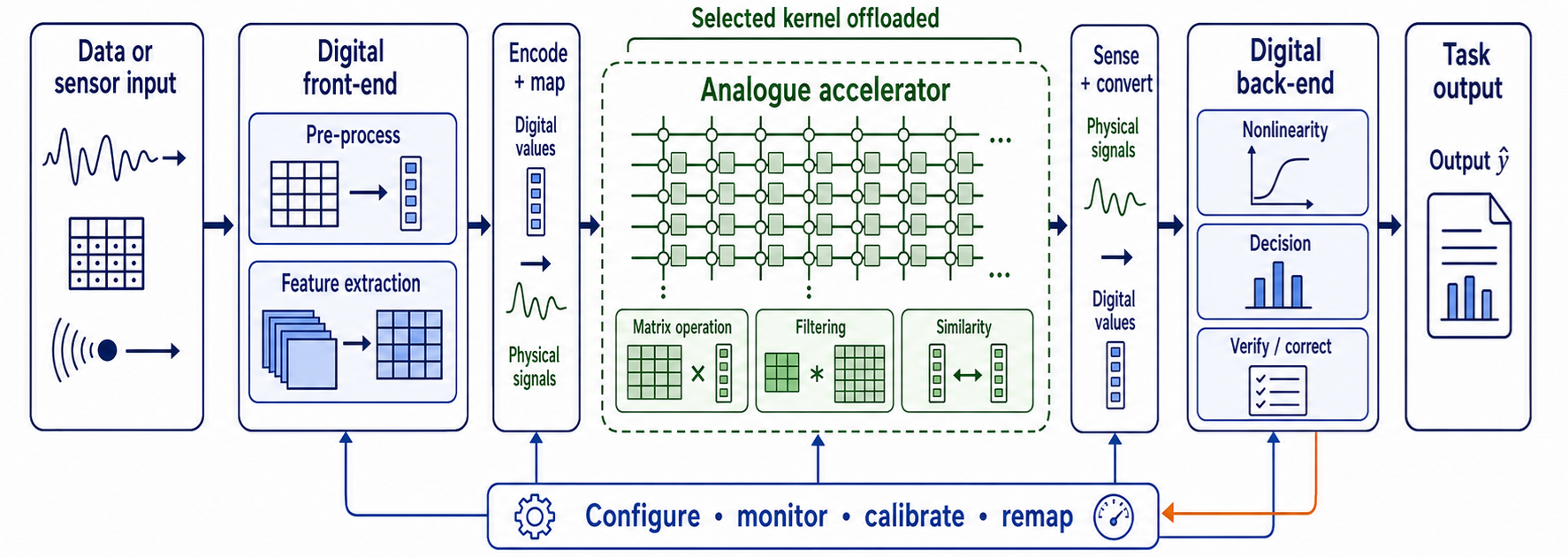}
    \caption{Analogue acceleration as one hybrid design pattern. A digital host
    prepares and schedules the workload, delegates suitable operations to an
    analogue accelerator and retains unsupported processing, orchestration and
    output verification.}
    \label{fig:hybrid1}
\end{figure*}

Recurring hybrid architectures can be expressed through five design patterns:

\begin{itemize}
    \item \textbf{Sensor-first front-end/back-end split:} a near-sensor,
    in-sensor, photonic or neuromorphic front end filters, transforms, compresses or detects events before digitisation or higher-level digital processing. A digital back end then classifies, decides, communicates or records the result~\cite{aspinity2024analogml}.

    \item \textbf{Controller/accelerator split:} a digital host prepares data,
    configures a physical accelerator, schedules suitable operations and
    retrieves the results. The digital processor retains orchestration,
    unsupported operations and system control
    ~\cite{sebastian2020memory,gallo2018mixed}. Figure~\ref{fig:hybrid1}
    illustrates this arrangement.

    \item \textbf{Coarse/fine split:} a physical engine produces a fast or
    approximate result, which a digital method refines through higher-precision
    computation or iterative correction~\cite{gallo2018mixed}.

    \item \textbf{Correction feedback:} digital logic estimates physical error,
    compensates for drift or variation and recalibrates or reprogrammes the
    physical engine when required~\cite{lottarini2021reaxion}.

    \item \textbf{Verification and fallback:} a specialised engine generates a
    candidate result, while digital logic checks critical outputs and repeats,
    corrects or redirects execution when confidence is insufficient.
\end{itemize}

\begin{figure*}[t]
    \centering
    \includegraphics[width=\textwidth]{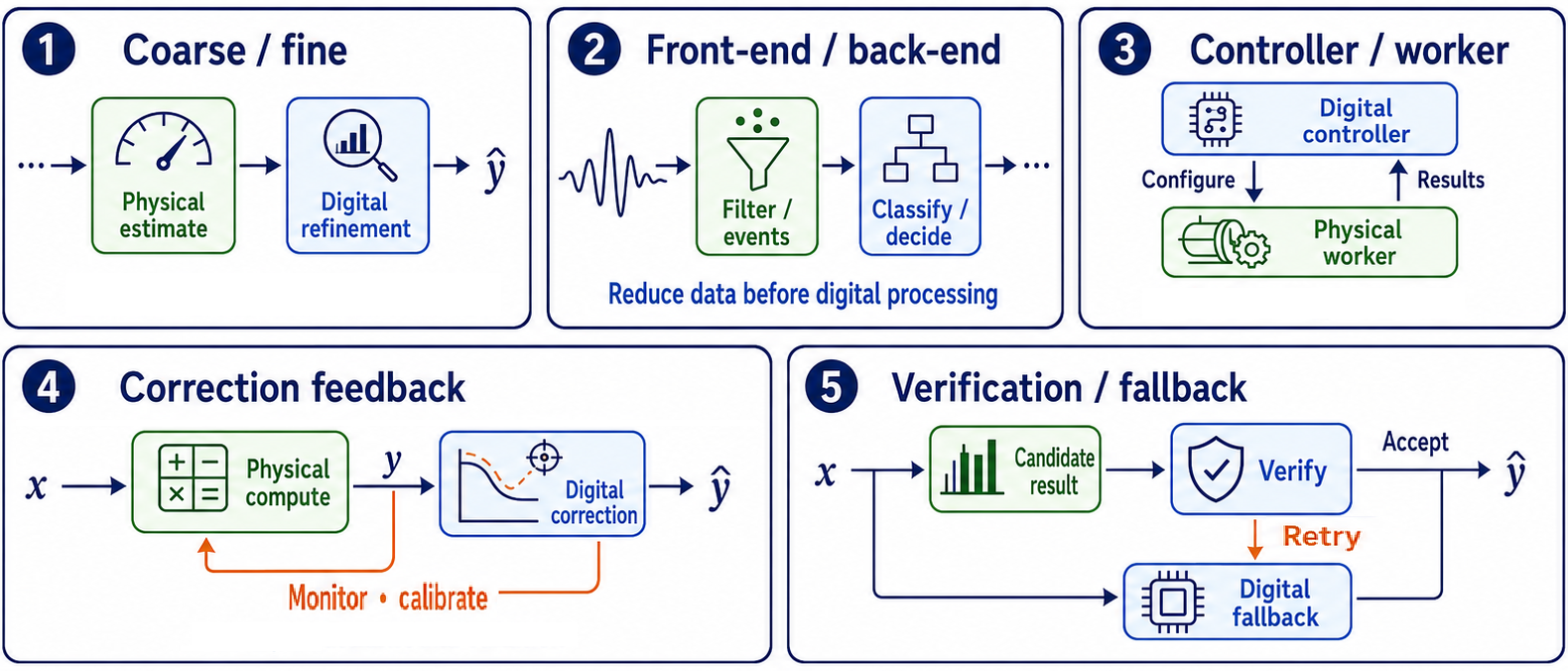}
    \caption{Common hybrid digital--analogue design patterns. Physical processing
    may occur before digitisation, operate as a digitally controlled
    accelerator, generate an approximate result for digital refinement, receive
    digital correction feedback or produce outputs that are digitally verified.}
    \label{fig:hybrid2}
\end{figure*}

Figure~\ref{fig:hybrid2} summarises these recurring forms of cooperation between
digital and physical processing~\cite{gallo2018mixed,sebastian2020memory,
lottarini2021reaxion}.

The patterns can be combined. For example, an analogue or event-driven front end may reduce sensor bandwidth before digitisation, an AIMC accelerator may execute a dense neural-network layer, and a digital controller may map the workload, apply nonlinear functions, manage calibration and verify the final decision.

\subsection{Deployment Criteria and Functional Complementarity}

A hybrid architecture is justified only when the workload matches the physical
engine and the advantage remains after system overheads are included. Four
conditions are particularly important:

\begin{itemize}
    \item \textbf{Workload compatibility:} the target operation should match the
    substrate, such as dense linear algebra for AIMC or photonics and sparse
    temporal processing for neuromorphic hardware.

    \item \textbf{Sufficient activity or reuse:} repeated operations should
    amortise programming, encoding, conversion and routing costs. For event-driven
    systems, useful sparsity must also exceed baseline, state-maintenance and
    routing overhead~\cite{indiveri2011neuromorphic}.

    \item \textbf{Acceptable accuracy and robustness:} task requirements must
    tolerate the effective precision and variability of the physical engine, or
    correction must recover the required accuracy without removing its energy or
    latency benefit.

    \item \textbf{End-to-end system advantage:} task energy or latency should
    improve relative to a strong digital baseline after data movement,
    interfaces, calibration, host processing and communication are counted
    ~\cite{horowitz2014computing,sebastian2020memory}.
\end{itemize}

Analogue and physical engines are strongest where continuous dynamics, spatial
parallelism or local transformations match the workload; digital systems
remain stronger for conditional control, exact arithmetic, secure execution
and mature software integration. Digital execution should remain the default when reproducibility is required, workloads change frequently, utilisation is low, or conversion, calibration and mapping costs outweigh the physical-core advantage. Hybrid computing is valuable only when this complementarity yields a measurable improvement at the level of the deployed task.

\section{Energy Efficiency as a System Property}
Energy efficiency is not unique to analogue computing. Digital systems have long used device scaling, power gating, dynamic voltage and frequency scaling (DVFS), near-threshold operation, heterogeneous cores, domain-specific accelerators, quantisation and sparsity to reduce energy consumption
~\cite{dally2020domain,frantar2023sparsegpt}. Mobile and embedded platforms combine these methods with memory-aware scheduling and duty cycling. Analogue, photonic and neuromorphic accelerators should therefore be compared with a strong, optimised digital baseline rather than with an unoptimised general-purpose implementation.

For edge and physical-AI systems, the energy boundary should include the whole task or mission. A suitable decomposition is

\begin{equation}
\begin{split}
E_{\mathrm{total}} ={}&
E_{\mathrm{sensing}}
+ E_{\mathrm{compute}}
+ E_{\mathrm{memory}}
+ E_{\mathrm{control}} \\
&+ E_{\mathrm{radio}}
+ E_{\mathrm{actuation}}
+ E_{\mathrm{idle}}
+ E_{\mathrm{calibration}} .
\end{split}
\label{eq:total_energy_physical_ai}
\end{equation}

Here, $E_{\mathrm{sensing}}$ includes sensor biasing, acquisition and signal
conditioning; $E_{\mathrm{compute}}$ includes digital processors and specialised accelerators; $E_{\mathrm{memory}}$ covers weight, activation and state movement across the memory hierarchy; and $E_{\mathrm{control}}$
includes scheduling, runtime management, security and logging. The terms $E_{\mathrm{radio}}$ and $E_{\mathrm{actuation}}$ apply when the system communicates or interacts physically with its environment. The idle term captures baseline and standby energy over the deployment interval, while
$E_{\mathrm{calibration}}$ includes testing, tuning, drift compensation and weight refresh. Calibration energy should be amortised over the number of tasks completed before recalibration is required.

Actuation may dominate energy consumption in drones, mobile robots and other mechanically active platforms, but this is workload and mission dependent. Compute savings may still matter when they extend duty cycle, reduce thermal load, or enable operation within a limited power budget. The value of a specialised accelerator should therefore be assessed through mission duration, task latency and completed work rather than through core-level arithmetic efficiency alone.

\section{Software, Compilers and Benchmarks}
The viability of hybrid digital--analogue systems depends on software that can
represent hardware constraints and coordinate heterogeneous execution. The
required stack includes device and circuit models, graph partitioning,
hardware-aware compilation, runtime scheduling, calibration support and
hardware-in-the-loop testing. A single universal stack may not be practical,
but shared abstractions are needed for portability and reproducibility.
A compiler cannot always treat a physical core as a deterministic arithmetic
unit. AIMC mapping may need to account for weight precision, programming
variation, conductance drift and converter resolution~\cite{gallo2018mixed};
photonic compilation for loss, phase settings, wavelength allocation and
electronic--optical boundaries; and neuromorphic mapping for neuron dynamics,
temporal discretisation, fan-in and routing capacity. These constraints should
guide partitioning across digital and physical engines within the system
boundary defined in Equation~\eqref{eq:total_energy_physical_ai}.

A deployment-oriented software stack requires four functions:

\begin{itemize}
    \item \textbf{Heterogeneous partitioning:} assigns compatible operators to
    digital, AIMC, photonic or neuromorphic engines while accounting for
    interfaces, memory movement and unsupported operations.

    \item \textbf{Hardware-aware numerical optimisation:} uses quantisation-
    aware or noise-aware training, mixed precision and targeted fine-tuning to
    maintain task performance within the precision and variability of the
    deployed hardware.

    \item \textbf{Runtime orchestration:} manages data transfer, duty cycling,
    power-state transitions, asynchronous events, accelerator availability and
    digital fallback.

    \item \textbf{Hardware-in-the-loop calibration:} measures the behaviour of
    the physical system and incorporates observed variation into mapping,
    correction or retraining when required.
\end{itemize}

\subsection{System-Level Benchmarking}

Peak operations per watt is insufficient for comparing substrates that use different representations, precisions and operating conditions. It may also exclude conversion, control, host processing, cooling and communication. A system-level benchmark should therefore report a multi-metric bundle

\begin{equation}
\mathcal{B} =
\left\{
E_{\mathrm{task}},
L_{\mathrm{task}},
A_{\mathrm{task}},
M_{\mathrm{move}},
C_{\mathrm{cal}},
R_{\mathrm{field}},
P_{\mathrm{prog}},
C_{\mathrm{deploy}}
\right\} .
\label{eq:evaluation_bundle}
\end{equation}

The coordinates have the following meanings:
\begin{itemize}
    \item $E_{\mathrm{task}}$: task energy, including interfaces, control and host involvement within the stated boundary.
    \item $L_{\mathrm{task}}$: end-to-end latency, including transfer,
     correction and synchronisation.
    \item $A_{\mathrm{task}}$: task performance (accuracy, error, reward or stability) under declared conditions.
    \item $M_{\mathrm{move}}$: volume and location of data movement; bit counts are useful, but energy depends on where and how transfer occurs.
    \item $C_{\mathrm{cal}}$: calibration overhead, reported as amortised energy, time, frequency and unavailable system time.
    \item $R_{\mathrm{field}}$: performance variation across temperature, supply, ageing and device variation.
    \item $P_{\mathrm{prog}}$: programmability and retargeting effort --- supported operators, compilation and engineering intervention required.
    \item $C_{\mathrm{deploy}}$: deployment cost, including packaging, integration, tooling, testing and certification.
\end{itemize}

\begin{table}[htbp]
\centering
\small
\caption{Reported accelerator results and their stated measurement boundaries.
Values are not directly comparable because workloads, precision and included
system components differ.}
\label{tab:reported-metrics}

\begin{tabularx}{\textwidth}{
@{}
>{\raggedright\arraybackslash}p{2.5cm}
>{\raggedright\arraybackslash}p{2.3cm}
>{\raggedright\arraybackslash}p{3.0cm}
>{\raggedright\arraybackslash}X
>{\centering\arraybackslash}p{0.9cm}
@{}}
\toprule
\textbf{Technology} &
\textbf{Metric} &
\textbf{Reported value} &
\textbf{Measurement boundary stated in the source} &
\textbf{Ref.} \\
\midrule
AIMC (PCM, 34-tile chip, 14\,nm) &
Energy efficiency &
Up to 12.4\,TOPS/W &
Chip-sustained result including inter-tile communication and analogue
peripheral circuitry, but excluding the auxiliary digital compute and SRAM
required by a complete product &
\cite{ambrogio2023analog} \\
\addlinespace
AIMC (PCM, 64-core chip, 14\,nm) &
Throughput and energy efficiency &
63.1\,TOPS, 9.76\,TOPS/W (one-phase, low precision);
16.1\,TOPS, 2.48\,TOPS/W (four-phase, high precision);
8-bit I/O &
Chip-level result including on-chip digital activation functions and the
inter-core communication network &
\cite{legallo2023mixed} \\

\addlinespace
Photonic accelerator (\(>16{,}000\) components) &
Cycle latency & Operation up to 1\,GHz; latency reported as potentially as low as
3\,ns per cycle & Optical multiply--accumulate core co-integrated with an electronic logic, memory and control chip through 2.5D packaging & \cite{hua2025integrated} \\

\bottomrule
\end{tabularx}
\end{table}

The bundle in Equation (21) is not a scalar score: its coordinates have different units and should not be collapsed into arbitrary weights. Comparisons should instead report the workload, task-performance constraint, batch size, duty cycle, process technology, memory hierarchy, converter precision, calibration policy, host involvement and measurement boundary, and should distinguish at least three results,(i) the specialised core, (ii) the complete hybrid system and
(iii) an optimised digital baseline at matched task performance and representative operating conditions. Benchmarking frameworks such as NeuroBench offer a useful basis for separating algorithm characteristics from physical-system measurements~\cite{yik2025neurobench}. Table~\ref{tab:reported-metrics} illustrates why this matters: each source states its own measurement boundary explicitly, and these boundaries differ across studies, so the values cannot be used to rank the underlying
technologies against one another.

\section{Bringing It Together: Pathways to Adoption}

Digital computing remains the default platform because it combines reliable hardware, mature software, established security mechanisms and proven deployment practices. Analogue, photonic and neuromorphic technologies are therefore likely to be adopted as specialised engines within digitally governed systems. Digital hosts will typically manage workload allocation, memory, calibration, verification and fallback when a physical result is uncertain or unreliable.

Recent chip-scale demonstrations show that these technologies are moving beyond isolated device experiments~\cite{legallo2023mixed,ambrogio2023analog,
hua2025integrated}. Their value, however, depends on whether an advantage remains after conversion, communication, control and calibration are included. A physical accelerator is justified when it improves task-level energy, latency or capability relative to an optimised digital baseline. The discussion also extends to emerging heterogeneous and distributed architectures, including systems that coordinate several specialised computing engines or support collaborative learning across devices.

Adoption is likely to be incremental and workload-specific. Promising
opportunities include repeated matrix operations, sparse temporal processing and always-on sensing, particularly in energy-constrained systems such as wearables, robots and drones. Progress also depends on connecting semiconductor materials, device and chip design with AI, algorithms and software. Materials and fabrication processes influence density, precision, variability, endurance and operating stability, while chip architecture determines how computing elements, memory, interconnects and mixed-signal interfaces are organised.
Early co-design across these levels can align physical capabilities with AI workloads, software tools and deployment requirements
~\cite{dally2020domain,sebastian2020memory,shekhar2024silicon}.

This transition also depends on development environments that make physical hardware accessible to AI and software researchers. Development-ready platforms provide documented operators, numerical constraints, calibration controls, performance counters and stable software interfaces. Device-aware simulators and emulators allow models to be trained, mapped and tested before physical hardware is widely available, while representing realistic precision, noise, drift, latency and conversion costs. Common intermediate representations, compiler support, software development kits and remotely accessible testbeds reduce barriers to experimentation and improve reproducibility
~\cite{rasch2023hardware,pedersen2024nir}. Comparison with measurements from physical hardware reveals whether simulation models accurately represent device, interface and system costs. Knowledge of analogue and physical computing becomes more useful when expressed through accessible concepts, models and tools that AI, digital-computing and software researchers can understand and apply without losing essential technical accuracy. A system-level co-design approach can prevent technically efficient devices from being developed around operations that remain difficult to map, integrate or deploy. This foundation is particularly relevant to heterogeneous, context-aware physical AI, where specialised computing engines, sensors, communication systems and actuators cooperate under changing
workloads, environments and resource constraints.

\section*{Acknowledgements}
This work was supported by the Engineering and Physical Sciences Research Council (EPSRC) through the TinyML UK Network (project reference UKRI3297). The TinyML-UK Network aims to support interdisciplinary collaboration across AI hardware, edge computing, neuromorphic computing, software, system tools and custom-built AI hardware. The authors also acknowledge the support of the EPSRC ProSensing: Low-Power, High-Speed, Adaptable Processing-In-Sensing Capability Project (Grant No. EP/Y030176/1).

\bibliographystyle{IEEEtran}
\bibliography{references}

\end{document}